\documentstyle[11pt]{article}
\makeatletter

\if@twoside
\else
\fi
\ifcase \@ptsize
\or 
\or 
\fi

\def\@citex[#1]#2{%
\if@filesw \immediate \write \@auxout {\string \citation {#2}}\fi
\@tempcntb\m@ne \let\@h@ld\relax \def\@citea{}%
\@cite{%
 \@for \@citeb:=#2\do {%
 \@ifundefined {b@\@citeb}%
 {\@h@ld\@citea\@tempcntb\m@ne{\bf ?}%
  \@warning {Citation `\@citeb ' on page \thepage \space undefined}}%
 {\@tempcnta\@tempcntb \advance\@tempcnta\@ne%
 \@tempcntb\number\csname b@\@citeb \endcsname \relax%
 \ifnum\@tempcnta=\@tempcntb %
 \ifx\@h@ld\relax%
 \edef \@h@ld{\@citea\csname b@\@citeb\endcsname}%
 \else%
 \edef\@h@ld{\ifmmode{-}\else--\fi\csname b@\@citeb\endcsname}%
 \fi%
 \else
 \@h@ld\@citea\csname b@\@citeb \endcsname%
        \let\@h@ld\relax%
      \fi}%
    \def\@citea{,\penalty\@highpenalty\,}%
  }\@h@ld
}{#1}}
\@addtoreset{equation}{section}
\def\section{\@startsection {section}{1}{\z@}{-3.5ex plus -1ex minus
 -.2ex}{2.3ex plus .2ex}{\large\bf\centering}}
\def\subsection{\@startsection{subsection}{2}{\z@}{-3.25ex plus%
 -1ex minus -.2ex}{1.5ex plus .2ex}{\sc}}
\gdef\@publabel{\hfil}
\gdef\@pubdate{\null}
\gdef\@pubnumber{\null}
\gdef\@author{\null}
\gdef\@title{\null}
\gdef\@abstract{\null}
\long\def\pubdate#1{\gdef\@pubdate{#1}}
\long\def\pubnumber#1{\gdef\@pubnumber{#1}}
\long\def\publabel#1{\gdef\@publabel{#1}}
\long\def\author#1{\gdef\@author{#1}}
\long\def\title#1{\gdef\@title{#1}}
\long\def\abstract#1{\gdef\@abstract{#1}}
\def\titlerelax{
\let\maketitle\relax
\let\settitleparameters\relax
\let\consolidatetitle\relax
\let\inittitlepage\relax
\let\finishtitlepage\relax
\let\titlepagecontents\relax
\let\multithanks\relax
\let\titlebaselines\relax
\let\@makepub\relax
\let\@maketitle\relax
\let\@makeauthor\relax
\let\@makeabstract\relax
\let\@maketitlenote\relax
\let\thanks\relax
\let\titlerelax\relax}
\def\titleclean
{\gdef\@titlenote{}
\gdef\@abstract{}
\gdef\@author{}
\gdef\@title{}
\gdef\@pubdate{}\gdef\@pubnumber{}\gdef\@publabel{}
\gdef\@dpublabel{}
}

\def\@makepub{\vbox to \z@{\hbox to \textwidth{\hfill
\@publabel \hfill
\llap{\parbox[t]{0.33\textwidth}{\raggedleft\@pubnumber}}}%
\vss}}
\def\@maketitle{\vskip 60pt \begin{center}
 {\LARGE \@title \par}
 \end{center}}

\def\@makeauthor{{%
\def\and{\smallskip {\normalsize \rm and\smallskip }}
\def\And{\medskip {\normalsize \rm and\\}\medskip}
\long\def\address##1{{\def\and{\\and\\}\medskip
				{\small \it \\##1\\}
}}
{\centering
 \vskip 3em
 \large \lineskip .75em
 \@author}
 \par}}
\def\@makedate{\vskip 1.5em
 {\raggedright \small \noindent\@pubdate \par}}
\def\@makeabstract{\vskip 1.5em
{\small
\begin{center}
{\bf ABSTRACT\vspace{-.5em}\vspace{0pt}}
\end{center}
\quotation \@abstract \endquotation}}
\def\maketitle{\titlepage
\let\footnotesize\small \setcounter{page}{0}
\@makepub
\vfil
\@maketitle
\@makeauthor
\vfil
\@makeabstract
\@thanks
\vfil
\@makedate
\if@restonecol\twocolumn \else \eject \fi
\titlerelax \titleclean
\setcounter{footnote}{0}
}

\newcommand{\ncm}{\newcommand}
\ncm{\cp}{ \!{}^+\!C}
\ncm{\cm}{ \!{}^-\!C}
\ncm{\ap}{ {}^+\!a}
\ncm{\am}{ {}^-\!a}
\ncm{\W}{{\cal W\! L}_2}
\ncm{\w}{\overline{\cal W\! L}_2}
\ncm{\jb}{\bar{\jmath}}
\ncm{\ub}{\bar{u}}
\ncm{\ul}{\underline}
\ncm{\gf}{{\bf g}}

\newtheorem{conjecture}{Conjecture}

\ncm{\be}{\begin{equation}
\addtolength{\abovedisplayskip}{\extraspaces}
\addtolength{\belowdisplayskip}{\extraspaces}
\addtolength{\abovedisplayshortskip}{\extraspace}
\addtolength{\belowdisplayshortskip}{\extraspace}}
\ncm{\ee}{\end{equation}}
\ncm{\bea}{\begin{eqnarray}
\addtolength{\abovedisplayskip}{\extraspaces}
\addtolength{\belowdisplayskip}{\extraspaces}
\addtolength{\abovedisplayshortskip}{\extraspace}
\addtolength{\belowdisplayshortskip}{\extraspace}}
\ncm{\eea}{\end{eqnarray}}
\ncm{\beas}{\begin{eqnarray*}
\addtolength{\abovedisplayskip}{\extraspaces}
\addtolength{\belowdisplayskip}{\extraspaces}
\addtolength{\abovedisplayshortskip}{\extraspace}
\addtolength{\belowdisplayshortskip}{\extraspace}}
\ncm{\eeas}{\end{eqnarray*}}
\newcommand{\ie}{{\it i.e}.\ }
\newcommand{\eg}{{\it e.g}.\ }
\newlength{\extraspace}
\newlength{\extraspaces}
\begin{document}
\pubnumber{DAMTP-93-66 \\UCLA/93/TEP/46 \\ hep-th/9312016 \\December 2, 1993}
\pubdate{}
\title{The Kazhdan-Lusztig conjecture for finite W-algebras}

\author{K.~DE VOS$^{1,2}$
\address{
Department of Applied Mathematics and Theoretical Physics\\
University of Cambridge,
Silver Street, Cambridge, CB3 9EW,
U.K.}
\And
P.~VAN DRIEL$^{3}$
\address{
Department of Physics\\
University of California at Los Angeles,
Los Angeles, CA 90024-1547, USA}
}

\footnotetext[1]{
Email: {\tt K.de\_Vos@damtp.cambridge.ac.uk}
}
\footnotetext[2]{
Supported by the U.K.\ Science and Engineering Research Council
}
\footnotetext[3]{
Email: {\tt vandriel@physics.ucla.edu}
}

\abstract{
We study the representation theory of finite W-algebras. After
introducing parabolic subalgebras to describe the structure of
W-algebras, we define the Verma modules and give a conjecture for the
Kac determinant. This allows us to find the completely degenerate
representations of the finite W-algebras. To extract the irreducible
representations we analyse the structure of singular and subsingular
vectors, and find that for W-algebras, in general the maximal
submodule of a Verma module is not generated by singular vectors only.
Surprisingly, the role of the (sub)singular vectors can be
encapsulated in terms of a `dual' analogue of the Kazhdan-Lusztig
theorem for simple Lie algebras. These involve dual relative
Kazhdan-Lusztig polynomials. We support our conjectures with some
examples, and briefly discuss applications and the generalisation to
infinite W-algebras.
}
\maketitle
\section{Introduction}
Since the introduction of W-algebras in conformal field theories by
Zamolodchikov, there has been a tremendous effort to somehow classify a
reasonable set of such algebras (see~\cite{w} for a review and further
references). In that context it has been a particularly fruitful
observation that a large class of W-algebras
can be obtained from affine Kac-Moody algebras by hamiltonian reduction.
Much of the power of this method resides in the fact that the inequivalent
reductions are in one to one correspondence with a simple Lie algebraic
structure: the inequivalent embeddings of ${\bf sl}_2$ into $\gf$ \cite{btv}.
This also allows for a complete construction of the quantum W-algebras by a
BRST procedure~\cite{ff,tb1}.\\
On the other hand, relatively little is known about the representation
theory of W-algebras, which in principle determines \eg correlation
functions  and critical exponents of models with W-symmetry.
The BRST method can be applied to study representations, by computing
the cohomology of representations of the underlying Kac-Moody
algebra~\cite{fkw}. However, this is a complicated problem, for which
the general solution is still lacking.\\
In this paper we report on some progress in a direct approach: the
analysis of the Verma modules of W-algebras. For simplicity we have
first restricted to finite W-algebras. These algebras have recently been
introduced as hamiltonian reductions of semisimple Lie algebras~\cite{w32,tb2}.
They correspond to the zero mode algebra (which closes on the
vacuum) of the corresponding infinite W-algebra. In this sense the
representation theories of both types of algebras are also intimately
linked.\\
Our main result is an explicit characterformula for irreducible completely
degenerate representations of the finite W-algebras, in terms
of characters of Verma modules. The result is a generalisation of the
Kazhdan-Lusztig (KL) theorems for regular integral highest weight
representations of simple Lie algebras. In that case, the structure of Verma
modules of $\gf$ is governed by the Weyl group: the Weyl orbit of the highest
weight predicts the location of singular vectors, and the Bruhat order of the
Weyl group gives the embedding pattern~\cite{bgg}. The
characterformula involves an alternating sum over the orbit, with
coefficients given by the KL polynomials~\cite{kl}.\\
In the case of finite W-algebras, remarkably all that happens is that the
Weyl group is replaced by a poset and the KL polynomials are
replaced by {\em dual relative} KL polynomials. The poset
corresponds to the coset of the Weyl group of $\gf$ over the Weyl group
of $\gf_S$, a minimal regular subalgebra that contains the embedded
${\bf sl}_2$ principally. These posets also describe the structure of
{\em generalised Verma modules}~\cite{lep} of $\gf$, based on the
parabolic subalgebra associated to $\gf_S$. Each poset gives rise to
{\em two} sets of relative KL polynomials~\cite{deo1}, $\{P(x)\}$ and
$\{\tilde{P}(x)\}$, which are roughly speaking each others inverse.
The `standard' set $\{P(x)\}$ gives the coefficients in the character
formula in the case of generalised Verma modules~\cite{caco}.
The `dual' set $\{\tilde{P}(x)\}$, which to our knowledge had not found any
such interpretation before, describes the coefficients in the case of Verma
modules of the finite W-algebras. \\
The outline of this letter is as follows. In section 2 the parabolic
subalgebra $\gf_S$ is employed to describe the Cartan subalgebra (CSA) and
roots of W-algebras. These notions are used in section 3 to define the
Verma modules. We give the Kac determinant that describes the location
of singular vectors. We define the completely degenerate representations,
and formulate our generalisation of the KL conjectures. In
section 4 two examples illustrate the results. We conclude with some comments
on the case of infinite W-algebras, and mention applications.
\section{Finite W-Algebras}
Let $\gf$ be an arbitrary simple Lie algebra and let $\delta$ be some grading
element which decomposes $\gf=\gf_- \oplus \gf_0 \oplus \gf_+$ in
negative, zero and positive eigenspaces. The W-algebra is obtained
from $\gf$ by imposing constraints on generators of $\gf_+$
\be \label{constraint}
\gf_+ - \chi(\gf_+) = 0,
\ee
where $\chi$ is some one-dimensional representation of $\gf_+$. We
will restrict to the case where $\delta$ is derived from some ${\bf
sl}_2$-embedding since in that case the resulting W-algebra has a
corresponding infinite dimensional analogue.\\
In the quantum version of this reduction~\cite{ff,tb1,tb2} one imposes the
constraint~(\ref{constraint}) using the BRST formalism. In the end, the
W-generators are expressed in terms of $\gf$-currents. Although the
BRST-construction of W-algebras is straightforward and algorithmic, the
structure of the W-algebra emerges only after one has done the explicit
calculations. Nonetheless: the W-algebra allows a triangular decomposition,
\ie a CSA and roots. To see this, we explore the connection between
${\bf sl}_2$ embeddings and parabolic subalgebras of $\gf$. \\
A parabolic subalgebra ${\bf p}$ of $\gf$ is a subalgebra which
contains the Borel subalgebra $\bf b$ of $\gf$: $ {\bf b} \subseteq {\bf p}
\subseteq {\bf g}$, it is fixed by specifying a subset of the set of
simple roots of ${\bf g}$. Let $\alpha_1,\ldots,\alpha_l$
be the simple roots of ${\bf g}$, and $S$ be any subset of $\{1,\ldots,l\}$.
This subset defines a semisimple subalgebra ${\bf g}_S$ of
${\bf g}$: it is generated by ${\bf h}_S$ ( $\subset {\bf h}$ the span
of the $h_i$ with $i \in S$) and the $g_{\pm \alpha_i}$ with $i \in
S$. The set of roots is $\Delta^S = \Delta \cap \amalg_{i\in S} {\bf
Z} \alpha_i$, and the set of positive roots is $\Delta_+^S = \Delta_+
\cap \Delta^S$. If we define the nilpotent subalgebras ${\bf n}_S =
\amalg_{\alpha \in \Delta^S_+} g_{\alpha}$, ${\bf n}_S^- =
\amalg_{\alpha \in \Delta^S_+} g_{-\alpha}$, then we have ${\bf g}_S =
{\bf n}_S^- \oplus {\bf h}_S \oplus {\bf n}_S$. The positive root
generators of ${\bf g}$ outside ${\bf g}_S$ form a subalgebra ${\bf u}
= \amalg_{\alpha \in \bar{\Delta}_+^S} g_{\alpha}$ (with
$\bar{\Delta}_+^S = \Delta_+ \setminus \Delta^S_+$), and
so do the negative root generators ${\bf u}^- = \amalg_{\alpha\in
\bar{\Delta}_+^S} g_{-\alpha}$. One now defines the
subalgebras
\be
{\bf r} = {\bf g}_S + {\bf h}, \qquad \qquad {\bf p}_S = {\bf r}
\oplus {\bf u}.
\ee
The subalgebra ${\bf r}$ is reductive in ${\bf g}$. ${\bf p}_S$ is the
parabolic subalgebra of ${\bf g}$ associated with $S$. ${\bf p}_S$ has
reductive part ${\bf r}$ and nilpotent part ${\bf u}$. One can use these
notions to define the associated decomposition
\be\label{trig}
{\bf g}={\bf u}^- \oplus {\bf r} \oplus {\bf u}.
\ee
The two trivial cases are $S=\emptyset$, where ${\bf p}_S={\bf b}$,
and $S=\{1,\ldots,l\}$, whence ${\bf p}_S = {\bf g}$.\\[5mm]
The relevance of this for the problem at hand is, that every
${\bf sl}_2$ embedding in $\gf$ is principally embedded in
a semisimple regular subalgebra $\gf_S$ for some\footnote{
$S$ is in general not uniquely determined by the embedding. Also
for some $\gf$ this correspondence is not complete.} $S$.
The branching of $\gf$ into ${\bf sl}_2$ irreps can be done in two
steps now. First consider
the branching with respect to $\gf_S$. By construction,
the nilpotent subalgebras ${\bf u}$ and ${\bf u}^-$ will each decompose
into irreps of $\gf_S$. It is easy to find these irreps: define an
equivalence relation on roots of ${\bf u}$, such that for $\alpha,\beta \in
\bar{\Delta}^S_+$ we have $\alpha \sim \beta \Leftrightarrow
\alpha-\beta \in \Delta^S .$ The equivalence classes in ${\bf u}$
correspond to the irreducible representations of $\gf_S$ (and
analogous for ${\bf u}^-$). We label these irreps by $\bar{\alpha}$
(and $-\bar{\alpha}$ for ${\bf u}^-$). These labels correspond to the
eigenvalues with respect to the centre of ${\bf r}$, denoted by ${\bf
\bar{h}}_S$, such that ${\bf h} = {\bf h}_S \oplus {\bf \bar{h}}_S$ and
\be
{\bf r} = {\bf g}_S \oplus {\bf \bar{h}}_S .
\ee
For each label $\bar{\alpha}$ there is a label for the associated irrep
of $\gf_S$, say $\lambda$, and we have the decomposition
\be
\label{gdeco}
{\bf g} = \left(\bigoplus_{\lambda,\bar{\alpha}>0} {\bf
u}^-_{\lambda,-\bar{\alpha}} \right) \oplus
({\bf g}_S \oplus {\bf \bar{h}}_S) \oplus
\left( \bigoplus_{\lambda,\bar{\alpha}>0} {\bf u}_{\lambda,\bar{\alpha}}
\right)
\ee
In the second step, irreps of $\gf_S$ will break up into ${\bf sl}_2$ irreps,
each of which is associated with a W-generator. The generators divide
into three sets denoted by ${\bf W}_-$, ${\bf W}_0$ and ${\bf W}_+$
according to the decomposition~(\ref{gdeco}).\\
The generators in ${\bf W}_0$ form a rank$\gf$-dimensional (maximal)
abelian subalgebra (since the ${\bf sl}_2$ is principally embedded in each
simple factor of $\gf_S$, and ${\bf \bar{h}}_S$ centralises $\gf_S$).
Strictly speaking that does not make it into a CSA since not all
generators in ${\bf W}_0$ diagonalise the algebra. We find that there are
three types of generators: 1) $U_1$ like generators, which are the
generators of the CSA of ${\gf}$ that survived the reduction
(corresponding to ${\bf \bar{h}}_S$). These generators obviously
diagonalise the algebra (the eigenvalues are the charges $\bar{\alpha}$);
2) central generators, since these generators commute with all other
generators of the W-algebra (there is one such generator for each
casimir of the largest simple factor in $\gf_S$) these generators only
give rise to trivial roots; and 3) non-semisimple generators, the algebra
cannot be diagonalised with respect to the adjoint action of these
generators. So there are also no roots associated with these generators.\\
The W-generators in ${\bf W}_+$ (${\bf W}_-)$ are labelled by their
charges with respect to the $U_1$-like generators (so $\pm
\bar{\alpha}$).  We denote the generators by $X_{j,\pm\bar{\alpha}}^{\mu}$
where $j$ is the ${\bf sl}_2$ spin, $\bar{\alpha}$ the charge, and $\mu$
keeps track of degeneracies. We have the triangular decomposition
familiar from Lie algebra theory:
\be\label{triw}
{\bf W}={\bf W}_-\oplus{\bf W}_0\oplus{\bf W}_+.
\ee
We stress that in contrast to ordinary Lie algebra theory, the
rootspaces are now in general higher dimensional, due to the existence
of the non-semisimple generators in ${\bf W}_0$.
\section{Highest weight representations and Verma modules}
%
%
Given the triangular decomposition~(\ref{triw}) we can define highest
weight representations. By definition a highest weight representation
$V$ contains a vector $v$ (highest weight vector) with
the following properties: 1) the CSA acts diagonally on $v$; 2) the
positive root generators annihilate $v$; 3) $V$ is generated by the
action of the negative root generators.\\
Every highest weight representation is a quotient of a universal
highest weight representation, the Verma module. These modules,
denoted $M(\lambda)$, are freely generated by the negative root
generators from the highest weight vector (for the moment, $\lambda$
somehow specifies the eigenvalue of the CSA on the highest weight vector).
Naturally, one is interested in the irreducible highest weight
representations, $L(\lambda)$, which can be obtained from the Verma
modules by quotienting with respect to their maximal submodule. Therefore,
one has to investigate the reducibility of the Verma modules.\\
Let us briefly summarise the results for a simple Lie algebra~\cite{bgg}.
A Verma module is reducible iff it contains a singular vector, \ie a
vector that is annihilated by the positive root generators. The weight of a
singular vector is related to the highest weight $\lambda$
by a shifted Weyl reflection
\be\label{orbit}
w \cdot \lambda = w(\lambda+\rho)-\rho,
\ee
and vice versa: for all such weights in the Verma module there is at most one
singular vector. Apart from the singular vectors there may also be
{\em subsingular} vectors. Such vectors become singular only after one
mods out the submodule generated by the singular vectors. Subsingular
vectors can occur only at the same weights as the singular vectors,
and therefore, the Jordan-H\"older series of a Verma module $M(\lambda)$
(this is the decomposition of $M(\lambda)$ in irreducible components $L(\mu)$)
is given by a (finite) set of weigths $\mu$ of the form~(\ref{orbit})
which may
occur with a multiplicity due to possible subsingular vectors. This results in
a characterformula for the irreducible representations in terms of characters
of Verma modules. The coefficients in this formula are given by the
KL polynomials, which have a geometrical interpretation as the dimensions of
certain intersection cohomology groups associated to Schubert varieties.
Restricting to regular integral weights, there is precisely one anti-dominant
weight -say $\lambda$- in the Weyl orbit. Define $L_w =
L(w \cdot\lambda)$ and $M_w = M(w \cdot \lambda)$ for all $w \in W$, the
Weyl group of $\gf$. Then we quote the following result~\cite{kl}
\be
 \mbox{ch}\, L_w = \sum_{y \leq w} \epsilon_y \epsilon_w
P_{y,w}(1)\: \mbox{ch} \, M_y ,
\ee
where the $P_{y,w}(x)$ are the KL polynomials and $P_{y,w}(1)$
denotes the value of $P_{y,w}(x)$ at $x=1$, $\epsilon_w=(-1)^{l(w)}$
where $l(w)$ is the length of $w$.\\[5mm]
We will propose an analogous formula for irreducible, completely degenerate
representations of finite W-algebras.\\
The Verma modules of the finite W-algebras are generated from a highest
weight vector $|\lambda^{\mu}>$ which satisfies\footnote{We ignore the spin
label of the generators.}
\be
X_0^{\mu}  |\lambda^{\mu}> = \lambda^{\mu}
|\lambda^{\mu}>,\qquad
X_{\bar{\alpha}}^{\mu}  |\lambda^{\mu}> = 0.
\ee
Ordering the roots with multiplicities, a basis for the Verma module is given
by the states
\be
X_{-\bar{\alpha}_1}^{\mu_1} X_{-\bar{\alpha}_2}^{\mu_2} \ldots
X_{-\bar{\alpha}_n}^{\mu_n}
| \lambda^{\mu} > \quad \mbox{for} \quad  (\bar{\alpha}_1,\mu_1) \geq
(\bar{\alpha}_2,\mu_2) \geq \ldots \geq
(\bar{\alpha}_n,\mu_n)
\ee
The anti-involution $X_{\bar{\alpha}}^{\mu} \mapsto X_{-\bar{\alpha}}^{\mu} $
allows one to define an invariant bilinear form on the Verma module.
The determinant of this form is the Kac determinant, which contains
information about the reducibility of the Verma module in the usual
way.

\begin{conjecture}
Consider the finite W-algebra associated to the ${\bf sl}_2$ embedding
that is principal in $\gf_S\subseteq\gf$. The Kac determinant at position
$\bar{\beta}$ in the Verma module $M(\Lambda)$ is given by
\be\label{kacdet}
M_{\bar{\beta}}(\Lambda) =
\prod_{k>0}\prod_{\alpha \in \bar{\Delta}_+^S}
(<\Lambda+\rho,\alpha>-\frac{k}{2}<\alpha,\alpha>)^{P(\bar{\beta}-k\bar{\alpha})}
\ee
Here $P(\bar{\alpha})$ is the Kostant partition function for the restricted
roots $\bar{\alpha}$ with their multiplicities (which gives the
multiplicities of the W Verma module). The W-weights are parametrised
by a $\gf$-weight $\Lambda$ in terms of invariants of the Weyl group
$W_S$ of $\gf_S$.
\end{conjecture}
Let us give some explanation. Products of factors in~(\ref{kacdet})
corresponding to roots in the same irrep of ${\gf}_S$ are
invariant\footnote{Note that $w(\alpha) \sim \alpha$ for
$\alpha \in \bar{\Delta}^+_S$ and $w \in W_S$, and that if
$\alpha \sim \beta$, then $\beta = w(\alpha)$ for some $w \in W_S$.}
under $W_S$. Therefore $M_{\bar{\beta}}(\Lambda)$ is a $W_S$-invariant
polynomial, $M_{\bar{\beta}}(w \cdot \Lambda) = M_{\bar{\beta}}(\Lambda)$
for all $w \in W_S$.
This means that it can be expressed in terms of the fundamental
polynomial invariants of $W_S$. There are precisely rank $\gf$
independent invariants $W_S$, with degrees corresponding to
the exponents of $\gf_S$~\cite{hum}. Our claim is that these
invariants correspond one-to-one with the eigenvalues of the CSA
generators of the W-algebra, in such a way that (\ref{kacdet}) is the
Kac determinant, in completely factorised form. The degree one
invariants correspond to the generators from ${\bar{\bf h}}_S$.
Furthermore there is one central generator for each Casimir of the
largest simple factor of $\gf_S$. The rest of the invariants is
associated with the non-semisimple generators.\\
In principle, the explicit parametrisation of the W-weights in terms of
$\gf$-weights $\Lambda$ can be extracted from the BRST-construction,
which expresses the W-generators in terms of $\gf$-generators. However,
then one also needs to determine the highest weight vector which in
general is rather involved. A more practical way is to compute a
few determinants directly to fix the parametrisation.\\
Given conjecture 1 we can now define the completely degenerate
representations, which have a maximal number of vanishing factors
in the Kac determinant. They are labelled by dominant regular
integral weights of $\gf$ (up to reflections in $W_S$). \\[5mm]
The $W_S$ invariance of the W-weights and the form~(\ref{kacdet})
of the Kac determinant implies that the weights of the W-singular
vectors are on the orbit of the coset $W_S \setminus W$. This coset
inherits the Bruhat ordering of the Weyl group $W$, making it into a
poset~\cite{hum}. The embedding diagram of the W-singular vectors
follows from this ordering on the poset. \\
To find the irreducible representations of finite W-algebras we need
to settle the question of the subsingular vectors. In contrast to the case
of simple Lie algebras, these vectors do show up in the completely
degenerate representations. It follows from the Kac determinant that
the weight of these subsingular vectors coincide with the weight of a
singular
vector. For the characterformula this means that non-trivial coefficients
may appear, as in the case of the general KL conjecture for simple Lie
algebras. By studying some explicit examples we have found that these
multiplicities are given by {\sl dual} KL polynomials. This is formulated
in the following:
\begin{conjecture}
The KL theorem for finite W-algebras is given by
\be\label{kar}
\mbox{\em ch} \, L^S_{\tau} = \sum_{\sigma\leq\tau \in W_S \setminus W}
\epsilon_{\sigma} \epsilon_{\tau} \tilde{P}^S_{\sigma,\tau}(1) \:
\mbox{\em ch} \, M^S_{\sigma},
\ee
where the $\tilde{P}^S_{\sigma,\tau}(x)$ are the dual relative
KL polynomials associated to the subgroup $W_S$.
\end{conjecture}
Here $M^S_{\tau}$ and $L^S_{\tau}$ respectively denote a Verma module
and an irreducible module for the W-algebra associated to the regular
subalgebra $\gf_S$. They are now labelled by an element of the poset
$W_S \setminus W$. We recall that these posets play a role in the theory
of generalised Verma modules, analogous to the full Weyl group for
the Verma modules~\cite{lep}. Generalised Verma modules are certain
quotients of Verma modules, defined when the highest weight is dominant
integral with respect to some regular subalgebra $\gf_S$ of $\gf$. There
is a `relative' KL conjecture~\cite{caco}, which expresses
the character of the irreducible representations in terms of characters
of generalised Verma modules. The coefficients are
given in terms of `standard' relative KL polynomials $P^S$, which are defined
for every such quotient of Coxeter groups~\cite{deo1}. There is a
second set of relative KL polynomials $\tilde{P}^S$, called `dual' since
they are the inverse of the standard relative KL polynomials in some
sense. The relative KL polynomials are related to the standard KL
polynomials~\cite{kl} by
\be
P^S_{\sigma,\tau} =  P_{w_Sx,w_Sy},\qquad
\tilde{P}^S_{\sigma,\tau} = \sum_{w \in W_S} \epsilon_w P_{wx,y}
\ee
where $x$ and $y$ are the minimal representatives of $\sigma$ and
$\tau$ respectively and $w_S$ is the longest element in $W_S$.
Remarkably, the dual polynomials appear in the above conjecture
for the W-algebras.
\section{Examples}
We have explicitly verified the results of the preceding section for
all finite W-algebras from ${\bf sl}_2,{\bf sl}_3$ and ${\bf sl}_4$.
We discuss two examples here. The main focus is on the structure of the
W-algebra and the representations. For details on the construction of
the algebra we refer to~\cite{tb2}, the calculation of the relative
KL-polynomials can be found in~\cite{dict}.
\subsection*{$\ul{3}\rightarrow \ul{2} + \ul{1} $}
Consider the finite W-algebra associated with the ${\bf sl}_2$
embedding that is principal in $\gf_S={\bf sl}_2\subset {\bf sl}_3$.
The parabolic subalgebra is fixed by choosing the ${\bf sl}_2$
associated with the root $\alpha_1$ (\ie $S=\{1\}$). The W-algebra
corresponding to this embedding is a polynomial deformation of
${\bf sl}_2$, which was discussed in~\cite{w32}. We first
summarise their results. There are 4 generators $\{e,f,h,w_2\}$
with $w_2$ a central element and
\be \label{fh}
[h,e]=2e,\quad [h,f]=-2f, \quad [e,f]=h^2 + w_2.
\ee
On a Verma module with highest weights $(h,w_2)$ the Kac determinant at
weight $(h-2p,w_2)$ reads
\be\label{kdw32}
\prod_{k=1}^{p} (w_2 + (h+1-k)^2 + \frac{1}{3}(k^2-1) )
\ee
The completely degenerate representations occur for
\be\label{par1}
h=\frac{2}{3}(p+k)-1,\qquad
w_2=-\frac{4}{9}(p^2-pk+k^2)-\frac{1}{3},
\ee
where $p$ is an arbirary positive integer and $k$ is a positive integer
smaller than $p$. These representations consist of reducible
$p$-dimensional representations with an irreducible $k$-dimensional
quotient.\\
To rederive these results in our framework we need the parametrisation
of the W-weights $h,w_2$ in terms of invariants of the Weyl group
$Z_2$ of $\gf_S$ which is generated by $w_{\alpha_1}$. Using the
explicit expressions for the W-generators in terms of the ${\bf sl}_3$
currents from the BRST-construction we find
\be\label{par2}
h=\frac{1}{3}(2\Lambda_1+4\Lambda_2+3) ,\qquad
w_2=-\frac{4}{3}<\Lambda,\Lambda+2\rho>-1.
\ee
With this parametrisation, the Kac determinant~(\ref{kdw32}) factorises
as
\be\label{kdfac}
\prod_{k=1}^p (\Lambda_2+1-k)(\Lambda_1+\Lambda_2+2-k),
\ee
which is in agreement with~(\ref{kacdet}) since $\bar{\Delta}_+^S$ consists
of the roots $\alpha_2$ and $\alpha_1+\alpha_2$. \\
Also, one may now parametrise the completely degenerate
representations~(\ref{par1}) in
terms of $\Lambda$ using~(\ref{par2}), it is easy to verify that this
gives rise to two sets of solutions
\be\label{par3}
\left\{\begin{array}{l}\Lambda_1=p-k-1\\ \Lambda_2=k-1
\end{array}\right.,\qquad
\left\{\begin{array}{l}\Lambda_1=k-p+1\\ \Lambda_2=p-1 \end{array}\right. .
\ee
The first set corresponds exactly to the integral regular weights in the
dominant Weyl chamber, the second set consists of the weights in the
$w_{\alpha_1}$-reflected Weyl chamber.\\
To arrive at a characterformula we consider the Jordan-H\"older-decomposition
of the Verma module $M_{h,w_2}$. Fix the parametrisation~(\ref{par3}) of
the completely degenerate weights by taking $\Lambda$ in the dominant
Weyl chamber. Clearly $w_2$ is Weyl-invariant, but the Weyl-orbit of the
weight $h$ of the singular vector~(\ref{par2}) consists of three elements
in the coset $Z_2\setminus S_3$ (see figure 1).
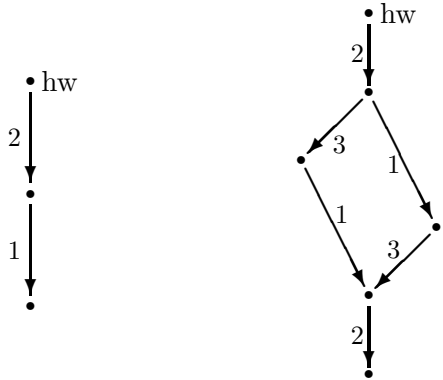
\begin{figure}[tbh]\begin{center}
\setlength{\unitlength}{0.3mm}
\begin{picture}(200,140)(0,30)
\thicklines
\put(0,50){\circle*{3}}
\put(0,100){\circle*{3}}
\put(0,150){\circle*{3}}\put(5,150){\makebox(0,0)[l]{\mbox{hw}}}
\put(0,145){\vector(0,-1){40}}
\put(0,95){\vector(0,-1){40}}
\put(-10,125){\makebox(0,0)[l]{\small $2$}}
\put(-10,75){\makebox(0,0)[l]{\small $1$}}
\put(150,20){\circle*{3}}
\put(150,55){\circle*{3}}
\put(150,145){\circle*{3}}
\put(150,180){\circle*{3}}\put(155,180){\makebox(0,0)[l]{\mbox{hw}}}
\put(180,85){\circle*{3}}
\put(120,115){\circle*{3}}
\put(142,37){\makebox(0,0)[l]{\small $2$}}
\put(158,75){\makebox(0,0)[l]{\small $3$}}
\put(135,91){\makebox(0,0)[l]{\small $1$}}
\put(142,162){\makebox(0,0)[l]{\small $2$}}
\put(158,113){\makebox(0,0)[l]{\small $1$}}
\put(140,122){\makebox(0,0)[r]{\small $3$}}
\put(150,175){\vector(0,-1){27}}
\put(150,50){\vector(0,-1){27}}
\put(152,141){\vector(1,-2){26}}
\put(122,111){\vector(1,-2){26}}
\put(147,142){\vector(-1,-1){24}}
\put(177,82){\vector(-1,-1){24}}
\end{picture}
\caption{\sl Embedding structure of the singular vectors in the completely
degenerate Verma module of W-algebras associated with the embedding $\ul{3}
\rightarrow\ul{2}+\ul{1}$ (left) and $\ul{4}\rightarrow\ul{2}+\ul{2}$ (right).
The numbers denote simple reflections, $\mbox{hw}$ is the highest weight. }
\end{center}
\end{figure}
Denoting the Verma modules by $M_{\sigma}$ with $\sigma=\{e,2,21\}$
and the irreducible representation at the same weight by $L_{\sigma}$,
as in the previous section, it is easy to see that
\be
\mbox{ch} M_{21}=\mbox{ch} L_{21} + \mbox{ch} L_2 + \mbox{ch} L_e,
\quad \mbox{ch} M_2=\mbox{ch} L_2 + \mbox{ch}L_e,
\quad \mbox{ch} M_e=\mbox{ch}L_e.
\ee
Inverting these relations we obtain
\be
\mbox{ch}L_{21}=\mbox{ch}M_{21}-\mbox{ch}M_2,
\quad \mbox{ch}L_2=\mbox{ch}M_2 - \mbox{ch}M_e,
\quad \mbox{ch}L_e=\mbox{ch}M_e.
\ee
{}From this one reads off the multiplicities $\tilde{P}_{\sigma,\tau}$
in eq.~(\ref{kar}), and they indeed coincide with the KL-polynomials
of this example (see table~\ref{tab5}).
\begin{table}[hbt]
\begin{center}
\begin{tabular}{c|ccc}
$P$& e&2& 21 \\ \hline
 e & 1& 1& 1 \\
 2&  & 1& 1 \\
 21 &  &  & 1
\end{tabular}\hspace{1cm}
\begin{tabular}{c|ccc}
$\tilde{P}$& e & 2 & 21 \\  \hline
 e  & 1& 1& 0 \\
2   &  & 1& 1 \\
 21 &  &  & 1
\end{tabular}
\caption{\sl Relative  KL matrices for $S=\{1\}$ in ${\bf sl}_3$.\label{tab5}}
\end{center}
\end{table}

\subsection*{$\ul{4}\rightarrow\ul{2}+\ul{2}$}
Consider the W-algebra that corresponds to the ${\bf sl}_2$ embedding that is
principal in $\gf_S={\bf sl}_2\oplus {\bf sl}_2\subset {\bf sl}_4$,
with $S=\{1,3\}$. It has 7 generators $\{w_2,J^a,S^a\}$ where $a=0,+,-$. The
generator $w_2$ is central and
\be
[J^a,J^b] = f^{ab}_{\quad c} J^c, \label{com1} \quad
[J^a,S^b] = f^{ab}_{\quad c} S^c, \label{com2} \quad
[S^a,S^b] = (w_2 - C_2) f^{ab}_{\quad c} J^c,\label{coms}
\ee
with normalisation $f^{0+}_{\quad +}=-f^{0-}_{\quad -}=f^{+-}_{\quad 0}=1$.
So the $J^a$ form an ${\bf sl}_2$ subalgebra, and the $S^a$ are in the adjoint
representation, the nonlinearity is hidden in the last commutator
in~(\ref{coms}) where $C_2=2(J^0J^0+J^+J^-+J^-J^+)$ is the Casimir
of the ${\bf sl}_2$ subalgebra, and this term does not commute with
the $S^a$.\\
The CSA is generated by $\{J^0,S^0,w_2\}$, but $S^0$ does not
diagonalise the algebra so the weight lattice is one-dimensional with
multiplicities $P(i)=i+1$. Using~(\ref{kacdet}) we find that the Kac
determinant $M_{j-p}(j,s,w_2)$ at weight $j-p$ reads
\be \label{kd4}
\prod_{k=1}^{p}\left( s^2 - (w_2-j(j+2)-(j+1-k)^2)(j+1-k)^2 \right)^{p-k+1},
\ee
where we have used the parametrisation
\be
j=\frac{1}{2}(\Lambda_1+2\Lambda_2+\Lambda_3+2),\quad
s=\frac{1}{4}(\Lambda_1-\Lambda_3)(\Lambda_1+\Lambda_3+2),
\ee
and $w_2=<\Lambda,\Lambda+2\rho> + 4$. These weights are invariant under the
Weyl subgroup $Z_2\times Z_2$ generated by the reflections $\{1,3\}$, For a
dominant integral weight the orbit under the coset
$(Z_2\times Z_2)\setminus S_4$ is depicted in figure 1.\\
At first sight, this embedding pattern is puzzling: all singular
vectors are descendant of the singular vector with weight
$w_2\cdot\Lambda$. So if one divides out the submodule
generated by the singular vectors one obtains an infinite dimensional
representation, with character $\mbox{ch} L_{2132}=\mbox{ch} M_{2132}-
\mbox{ch} M_{213}$. However, as is signaled by the non-trivial
multiplicity in the KL-matrix in table~\ref{tab1}: there is a
subsingular vector with weight $w_{213}\cdot\Lambda$. Adding this
vector to the maximal submodule -taking into account the overlap of
the submodules- results in the correct finite dimensional irreducible
representation with character
\be
\mbox{ch} L_{2132}=\mbox{ch} M_{2132}-\mbox{ch} M_{213}-\mbox{ch}
M_2+\mbox{ch} M_e.
\ee
Again this result is reproduced correctly by~(\ref{kar}) using
the KL-polynomials of table~\ref{tab1}.
\begin{table}[bth]
\begin{center}
\begin{tabular}{c|cccccc}
$P$ & e&2& 21&23&213&2132\\ \hline
 e  & 1& 1& 1& 1& 2& 1 \\
 2  &  & 1& 1& 1& 1& 1 \\
 21  &  &  & 1&  & 1& 1 \\
 23 &  &  &  & 1& 1& 1 \\
213 &  &  &  &  & 1& 1 \\
2132&  &  &  &  &  & 1 \\
\end{tabular} \hspace{1cm}
\begin{tabular}{c|cccccc}
$\tilde{P}$ & e&2&21&23&213&2132\\ \hline
 e & 1& 1& 0& 0& 0& 1 \\
2  &  & 1& 1& 1& 1& 1 \\
21  &  &  & 1&  & 1& 0 \\
23  &  &  &  & 1& 1& 0 \\
213 &  &  &  &  & 1& 1 \\
2132 &  &  &  &  &  & 1 \\
\end{tabular}
\caption{\sl Relative KL matrices for $S=\{1,3\}$ in ${\bf sl_4}$.\label{tab1}}
\end{center}
\end{table}
\section{Discussion}
We have studied the representation theory of finite W-algebras. After
recognizing the relevance of parabolic subalgebras to the description
of the structure of W-algebras, we have defined the Verma modules and
given the Kac determinant~(\ref{kacdet}). This allowed us to find the
completely degenerate representations of the finite W-algebras. To
extract the irreducible representations we have analysed the structure
of singular and subsingular vectors, and found that for W-algebras, in
general the maximal submodule of a Verma module is not generated by
singular vectors only. Surprisingly, the role of the
(sub)singular vectors can be encapsulated in terms of a `dual'
analogue~(\ref{kar}) of the KL-theorem for simple Lie algebras, which
involves the dual relative KL polynomials. \\
Note that the correspondence between ${\bf sl_2}$ embeddings
and parabolic subalgebras may not be one-to-one. An ambiguity in the
choice of parabolic subalgebras leads to different types of Verma
modules for the same W-algebra (corresponding to a different choice of
positive W-roots), and different characterformulas. For
finite irreps, this results in some interesting character identities. \\
In all the examples we studied we were able to define some limit in
which the W-algebra reduces to the subalgebra $\gf_0 \subset \gf$, \ie
the W-algebra can be viewed as a deformation of $\gf_0$. The
deformation results in the truncation of $\gf_0$ irreps to W irreps.
This is not so surprising since the W-algebras are symmetry algebras
of non-abelian Toda models, \ie $\gf_0$ sigmamodels with a non-trivial
selfinteraction~\cite{w32}. The relation to the quantum Miura transformation
of~\cite{tb2} needs to be investigated. \\
For the W-algebra $4 \rightarrow 2+2$  there is a nice physical interpretation
of the truncation. The algebra~(\ref{coms}) is the dynamical symmetry algebra
of a particle moving in a 3-dimensional space with constant curvature
under the influence of a Coulomb potential~\cite{Higgs}.
For negative curvature, the Casimirs $D_1,D_2$ of the
algebra~(\ref{coms}) are subjected to the physical constraints
\be\label{hicon}
D_1 = -J.S =0 ,\qquad
D_2 = -S^2 - \frac{1}{2}C_2^2 + (w_2-2)C_2 = \nu^2,
\ee
where $\nu\equiv R/R_B$ measures the curvature~\cite{Higgs}.
The Hamiltonian $H=-w_2/2$ has a spectrum
$E_j=-(j(j+2)+\nu^2/(j+1)^2)/2$, as follows directly from~(\ref{hicon}).
For zero curvature, where the algebra reduces to the (linear)
Runge-Lenz algebra $O(4)$, all states with $E<0$ are bound states.
The non-zero negative curvature effectively truncates this bound spectrum
at the point where the energy levels start to cross. It is easily
verified that is exactly as predicted by the
Kac determinant~(\ref{kd4}).\\
Obviously, one would like to generalise these results to infinite
W-algebras, and in fact this seems rather straightforward, since all
the necessary ingredients already exist. The parametrisation of the W-weights
is fixed by the analysis of the underlying finite W-algebra,
and the relative KL polynomials are also defined, although they are no
longer related by an inversion relation. The explicit verification
will be a lot harder though, and a general proof of the conjectures is
lacking so far. Work on this is in progress. \\[5mm]
{\bf Acknowledgement}. KdV would like to thank G.\ Watts, A.\ Kent and
F.\ Malikov for discussions, and P.\ Bowcock for drawing our attention
to~\cite{Higgs}. PvD acknowledges discussions with K.\ Pilch and P.\
Bouwknegt.

\end{document}